\documentclass[reprint,amsmath,amssymb,accepted=2021-12-24,a4paper]{quantumarticle}
\pdfoutput=1

\usepackage[numbers,sort&compress]{natbib}

\usepackage[dvipsnames]{xcolor}

\newcommand{\correction}[1]{\textcolor{black}{#1}}

\newcommand{\Sset}{\mathcal{S}}
\newcommand{\Hset}{\mathcal{H}}
\newcommand{\nmax}{n_{\text{max}}}
\newcommand{\mmax}{m_{\text{max}}}
\newcommand{\edet}{\eta_{\text{det}}}
\newcommand{\mean}[1]{\langle #1 \rangle}

\newcommand{\U}{\mathcal{U}}
\newcommand{\V}{\mathcal{V}}
\newcommand{\W}{\mathcal{W}}

\newcommand{\A}{\mathsf{A}}
\newcommand{\X}{\mathsf{X}}
\newcommand{\B}{\mathsf{B}}
\newcommand{\Cs}{\mathsf{C}}
\newcommand{\N}{\mathsf{N}}
\newcommand{\M}{\mathsf{M}}
\newcommand{\ps}{\mathcal{G}^{(n)}}
\newcommand{\pinching}{\mathcal{T}}
\newcommand{\keep}{\mathtt{keep}}
\newcommand{\discard}{\mathtt{discard}}
\newcommand{\postselect}{\mathtt{PS}}

\newcommand{\E}{\mathcal{E}}
\newcommand{\T}{\tau}

\usepackage[utf8]{inputenc}
\usepackage[english]{babel}
\usepackage[T1]{fontenc}
\usepackage{amssymb}

\usepackage{array}
\usepackage{booktabs}
\usepackage{tikz}
\usepackage{pifont}
\newcommand{\cmark}{\ding{51}}%
\newcommand{\xmark}{\ding{55}}%

\usepackage{mathtools}
\usepackage{amsmath}
\usepackage{bm}
\usepackage{stmaryrd}
\usepackage{physics}
\usepackage{amsfonts}
\usepackage{subcaption}
\usepackage{dcolumn}
\usepackage{graphicx}
\usepackage[colorlinks=true, allcolors=blue]{hyperref}
\hypersetup{
  colorlinks   = true, 
  urlcolor     = blue, 
  linkcolor    = blue, 
  citecolor   = blue 
}
\usepackage[justification=justified]{caption} 
\usepackage{ulem}
\renewcommand{\emph}{\textit}
\usepackage{ragged2e}
\usepackage{dsfont}
\DeclareCaptionJustification{justified}{\justifying}
\makeatletter 

\newcommand{\ah}{\hat{a}}
\newcommand{\bh}{\hat{b}}

\newcommand{\qh}{\hat{Q}}
\newcommand{\1}{\mathds{1}}

\DeclareMathOperator{\sign}{sign}
\newcommand*\diff{\mathop{}\!\mathrm{d}}

\makeatother
\begin{document}

\title{Discrete-variable quantum key distribution with homodyne detection}

\author{Ignatius W. Primaatmaja}
\email{ign.william@gmail.com}
\affiliation{Centre for Quantum Technologies, National University of Singapore, 117543, Singapore}
\author{Cassey C. Liang}
\affiliation{Department of Electrical \& Computer Engineering, National University of Singapore, 117583, Singapore}
\author{Gong Zhang}
\affiliation{Department of Electrical \& Computer Engineering, National University of Singapore, 117583, Singapore}
\author{Jing Yan Haw}
\affiliation{Department of Electrical \& Computer Engineering, National University of Singapore, 117583, Singapore}
\author{Chao Wang}
\affiliation{Department of Electrical \& Computer Engineering, National University of Singapore, 117583, Singapore}
\author{Charles C.-W. Lim}
\email{charles.lim@nus.edu.sg}
\affiliation{Centre for Quantum Technologies, National University of Singapore, 117543, Singapore}
\affiliation{Department of Electrical \& Computer Engineering, National University of Singapore, 117583, Singapore}

\begin{abstract}
Most quantum key distribution (QKD) protocols can be classified as either a discrete-variable (DV) protocol or continuous-variable (CV) protocol, based on how classical information is being encoded \correction{and decoded}. We propose a protocol that combines the best of both worlds---the simplicity of quantum state preparation in DV-QKD together with the cost-effective and high-bandwidth of homodyne detectors used in CV-QKD. Our proposed protocol has two highly practical features: (1) it does not require the honest parties to share the same reference phase (as required in CV-QKD) and (2) the selection of decoding basis can be performed after measurement. We also prove the security of the proposed protocol in the asymptotic limit under the assumption of collective attacks. Our simulation suggests that the protocol is suitable for secure and high-speed practical key distribution over metropolitan distances.
\end{abstract}

\maketitle

\section{Introduction}
Quantum key distribution (QKD) provides an information-theoretic method to exchange secret keys between distant parties, whose security is promised by the laws of quantum mechanics~\cite{scarani_security_2009, xu_secure_2020, pirandola2020advances}. Based on how classical information is being encoded and decoded, QKD can be divided into two broad protocol categories, namely, discrete-variable (DV) protocols and continuous-variable (CV) protocols. In the former, the information is typically encoded into discrete optical modes of a single photon, e.g., polarisation or time bin. In this case, single-photon detectors are normally used to perform decoding. In the latter, quantum states are described in an optical domain where the eigenstates are continuous and have infinite dimension~\cite{lvovsky_continuous-variable_2009,weedbrook_gaussian_2012}, e.g., using Gaussian optical states. One of the key benefits of CV-QKD is the use of homodyne detectors, which possess appealing features like high quantum efficiency, cost-effectiveness, and room-temperature operation. Moreover, homodyne detectors can be readily integrated into a photonic integrated circuit~\cite{zhang_integrated_2019,raffaelli_homodyne_2018,tasker2020silicon,bruynsteen_integrated_2021}, which holds great promise for monolithic CMOS-compatible fabrication and large-scale integrated quantum networks. 

Like DV-QKD, the current research trend of CV-QKD is focused on closing the gaps between theory and experiment. One prominent example is the GG02 protocol proposed by Grosshans and Grangier~\cite{grosshans_quantum_2003}, which requires two key assumptions: (1) the users are able to perform ideal Gaussian modulation and that (2) their local oscillators (LOs) are coordinated/calibrated (in terms of relative phase and wavelength). While in theory these conditions are well defined and understood, their practical implementations are not straightforward. Indeed, in the case of the first assumption, one would need an infinite amount of randomness to simulate the required Gaussian distribution, which is clearly not possible in practice. To overcome this gap, one solution is to consider the discrete approximation of Gaussian modulation~\cite{kaur2021asymptotic}, or alternatively, discrete-modulated CV-QKD protocols based on constellations of coherent states (or displaced thermal states)~\cite{zhao2009asymptotic, bradler2018security,leverrier2009unconditional,ghorai2019asymptotic,lin2019asymptotic, lin_trusted_2020, upadhyaya_dimension_2021, matsuura2021finite, denys2021explicit, papanastasiou_quantum_2018, papanastasiou_continuous_2021}. Moreover, working with discrete modulation protocols has another advantage, in that could significantly reduce the implementation complexity and computational resources required by the classical post-processing layer~\cite{weedbrook_gaussian_2012,diamanti_distributing_2015,leverrier_multidimensional_2008,leverrier2009unconditional,ghorai2019asymptotic,lin2019asymptotic}.

\begin{table*}[t]
  \centering
  \resizebox{\textwidth}{!}{
  \begin{tabular}{*5c}
    \toprule
    Protocol class & modulation type & detector type  & pilot pulse & maximum\\
    &  & & requirement &  transmission distance~\cite{xu_secure_2020}\\
    \midrule
    Standard DV-QKD   &  discrete & single-photon detector   &  \xmark & long (>100km)\\
    Gaussian-modulated CV-QKD   &  continuous & homodyne/heterodyne  & \cmark & mid\\
    Discrete-modulated CV-QKD   &  discrete & homodyne/heterodyne  & \cmark & mid\\
    DV-QKD with homodyne detection   &  discrete & homodyne/heterodyne & \xmark & metropolitan (<20km)\\
    \bottomrule
    \end{tabular}}
    \caption{{\textbf{Typical features of different classes of QKD protocols.}}~Here, we briefly compare between standard DV-QKD protocols, Gaussian-modulated CV-QKD protocols, discrete-modulation CV-QKD protocols and DV-QKD protocols with homodyne detection. Some popular protocols of standard DV-QKD are BB84~\cite{bb84}, B92~\cite{b92}, SARG04~\cite{sarg}, as well as distributed-phase-reference protocols~\cite{cow,dps}. An example of Gaussian-modulated CV-QKD is the aforementioned GG02 protocol~\cite{grosshans_quantum_2003}, while some examples of discrete-modulated CV-QKD can be found in Refs.~\cite{zhao2009asymptotic,bradler2018security,leverrier2009unconditional,ghorai2019asymptotic,lin2019asymptotic, lin_trusted_2020, upadhyaya_dimension_2021, matsuura2021finite, denys2021explicit, papanastasiou_quantum_2018, papanastasiou_continuous_2021,kaur2021asymptotic}. Finally, DV-QKD with homodyne detection includes Ref.~\cite{qi_bb84_2021} and this work. Note that the header `pilot pulse requirement' refers to the need of sending pilot pulses in CV-QKD to generate the LO locally using independent laser sources~\cite{qi_generating_2015, soh_self-referenced_2015}. Here, the term `standard DV-QKD' refers to protocols that encode classical information into optical modes and decode them via single-photon detection. For CV-QKD, we restrict our consideration to protocols with a local LO to avoid side-channel attacks on the LO.}
    
    \label{tab: features}
\end{table*}

For the second assumption, one can try to distribute a common LO together with the quantum signals using time/polarisation-division multiplexing, an approach commonly known as the \emph{transmitted LO} scheme. However, this approach is not entirely secure as it has been shown that the transmitted LO's intensity can be manipulated to break the security of the protocol~\cite{ma_LOfluctuation_2013,jouguet_preventing_2013,huang_quantum_2014}. A good solution is to use the so-called \emph{local LO} scheme, where the LOs are prepared independently~\cite{qi_generating_2015,soh_self-referenced_2015,huang_high-speed_2015}. Here, the relative phase between signal and LO has to be tracked and \correction{corrected accordingly~\cite{soh_self-referenced_2015, huang_high-speed_2015,wang2018high,qi_generating_2015,kleis2017continuous,marie_self-coherent_2017, laudenbach2019pilot}}
, with the help of pilot pulses acting as a phase reference. As such, the scheme is immune against side-channel attacks on the transmitted LO. In addition, it could reduce the power requirement for transmitted LO and bypass the excess noises caused by the transmission of a strong LO (which was needed previously). However, the local LO scheme typically requires a carefully designed phase tracking system and \correction{phase correction or compensation system}, which may increase the experimental complexity of the protocol~\cite{huang_high-speed_2015,wang2018high,kleis2017continuous, laudenbach2019pilot, marie_self-coherent_2017}.

Given the pros and cons of DV-QKD and CV-QKD, it is thus of interest to study hybrid QKD schemes that tap on the best of both approaches. In particular, this raises the question whether one could combine discrete encoding with continuous decoding without a common phase reference, using the transmission of randomly prepared single photons like in decoy-state BB84 QKD. Indeed, using homodyne detection to decode quantum information encoded in single photons is not trivial---the inherent quantum noise of homodyne detection generally engulfs the presence of single photons in a single-shot setting. Consequently, this makes it very hard to distinguish the underlying quantum code words. However, not all is lost: one can still exploit the noisy photon counting features of homodyne detection to learn some information about the underlying photon number distribution, as was recently shown in Ref.~\cite{qi_bb84_2021} assuming the use of a perfect single-photon source for quantum encoding. 

Here, arising from an independent work based on the estimation of input-output photon number distribution over an untrusted photonic channel~\cite{lavie2021estimating}, we present a practical hybrid QKD protocol based on decoy-state method and homodyne detection. More specifically, in our proposal, the quantum transmitter uses a standard phase-randomised coherent laser source, while the quantum receiver uses a single homodyne detector whose local oscillator's phase is randomised. As such, unlike CV-QKD, there is no need to distribute a common phase reference between the transmitter and receiver. The other key feature of our protocol is that the inherent quantum noise of the homodyne detector can be isolated from the measurement statistics related to the security analysis. This allows us to effectively restrict the analysis to the untrusted quantum channel. Moreover, unlike standard BB84 QKD, our protocol does not require the receiver to select a random basis for each measurement---basis selection can be deferred to the post-processing stage after all the measurements are done. Concerning its security performance, using state-of-the-art security proof techniques, we find that the protocol is able to distribute secret keys over metropolitan distances (up to 15 km fibre length) based on typical device and channel parameters. For an overview of the different practical features of different classes of QKD protocols, we refer the readers to Table~\ref{tab: features}.

A more detailed description of our proposed protocol is presented in Section~\ref{sec: protocol}. Then, in Section~\ref{sec: security analysis}, we analyse the security of the protocol in the asymptotic limit under the assumption of collective attacks. Next, in Section~\ref{sec: simulation}, we simulate the performance of the protocol in realistic settings.

\section{Protocol} \label{sec: protocol}
As mentioned, in our proposed protocol, Alice prepares time-phase BB84 states \cite{bb84} using phase-randomised weak coherent pulses (with decoy-state method \cite{lo_decoy_2005}) and
Bob performs random quadrature measurement using homodyne detection.

For time-phase encoding implementation of the BB84 protocol, when Alice chooses the $Z$-basis, she sends a signal in either the `early' or `late' pulse depending on her randomly chosen bit value. In the language of quantum optics, these temporal modes could be associated with orthogonal annihilation operators: namely, $\ah_0$ for the `early' pulse and $\ah_1$ for the `late' pulse. On the other hand, when the $X$-basis is chosen, Alice sends two successive pulses that are either `in-phase' or `out-of-phase'. The corresponding annihilation operators associated to these modes are given by $\ah_{\pm} = (\ah_0 \pm \ah_1)/\sqrt{2}$. When implemented using weak coherent source, Alice would then prepare a coherent state (with its global phase randomised) associated to the appropriate annihilation operator.

\begin{figure}[t]
\centering
\includegraphics[width=1\columnwidth]{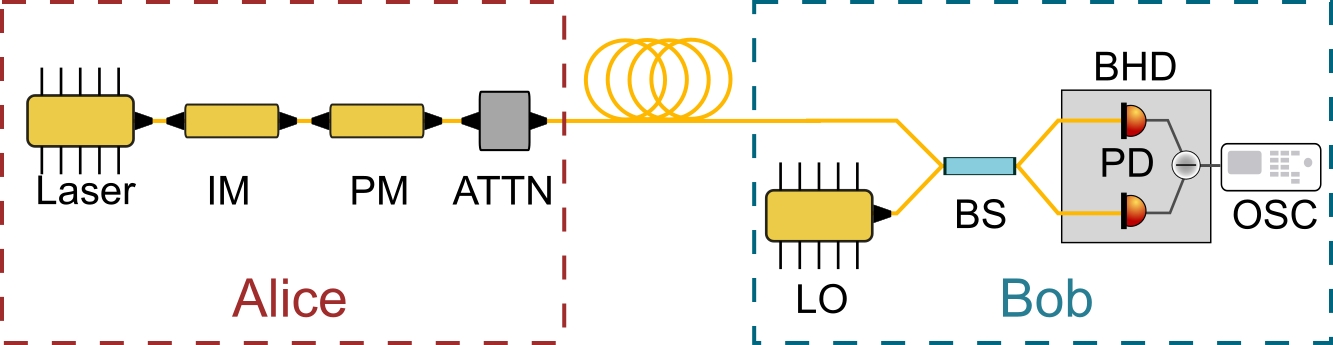}%
\vspace*{-1mm}
\caption{\label{exp_setup}\textbf{Schematic setup of the protocol}. Alice prepares time-phase BB84 states from phase-randomised coherent states using a set of intensity modulator (IM), phase modulator (PM) and variable optical attenuator (ATTN). After the quantum channel, Bob performed a phase-randomised quadrature measurement upon the input states using an independent local oscillator (LO) with a balanced homodyne detector (BHD). OSC: oscilloscope.}
\vspace*{-0.2cm}
\end{figure}

The schematic setup of the protocol is shown in Fig.~\ref{exp_setup}. The protocol runs as follows:
\begin{enumerate}
\item \textit{State preparation:} For each round, Alice randomly chooses an intensity setting $\mu \in \{\mu_0, \mu_1, \mu_2, \mu_3\}$ with their corresponding probability $p_{\mu_i}$ and a basis $\X \in \{Z,X\}$ with probability $\{p_Z, 1-p_Z\}$ respectively. In this protocol, keys are only generated when $\mu = \mu_0$ and $\X = Z$. Next, she chooses a bit value $\A \in \{0,1\}$ with uniform probability. Based on these choices, she would then prepare a phase-randomised coherent state with intensity $\mu$ in the appropriate optical mode. Finally, she sends the prepared states to Bob via an untrusted quantum channel.
\vspace{-0.2cm}
\item \textit{Measurement:} Bob performs homodyne measurement on the state that he receives using a local oscillator with a random global phase (the phase of the local oscillator for the `late' pulse is the same as the one for the `early' pulse). He records the outcome of the homodyne measurement for the `early' time-bin (denoted by $q_0$) and for the `late' time-bin (denoted by $q_1$). He also calculates $q_{\pm} = (q_0 \pm q_1)/\sqrt{2}$.

For each $\beta \in \{0,1,+,-\}$, he maps the continuous outcome $q_{\beta}$ into a discrete bin $\nu_\beta$. If the size of the bin is $\delta$, then for all $\nu \in \mathbb{Z}$, we denote the interval $[\nu \delta, (\nu+1) \delta)$ by $I_{\nu}$. Bob has $\nu_\beta = \nu$ if and only if $q_\beta \in I_{\nu}$.

\correction{(Remark: In practice, the outcome $q_{\beta}$ is already digitised. In this case, $\nu_\beta$ can be seen as an additional coarse-graining.)}

They repeat Step 1 and 2 for $N$ times.
\vspace{-0.2cm}
\item \textit{Basis and intensity announcement and decoding:} For all the rounds, Alice declares her basis choice $x$ and her intensity setting $\mu$. Finally, depending on Alice's basis choice, Bob assign his bit
value $\B = b$ according to the following decoding scheme:

\noindent For $Z$-basis:
$$
b=
\begin{cases}
0, &\abs{q_{0}} \geq \T, \,  \abs{q_{1}} < \T \\
1, &\abs{q_{0}} < \T, \, \abs{q_{1}} \geq \T \\
?, &\abs{q_{0}}, \abs{q_{1}} \geq \T \\
\varnothing, &\abs{q_{0}}, \abs{q_{1}} < \T
\end{cases}.
$$

\noindent For $X$-basis:
$$
b=\begin{cases}
0, & \left(\abs{q_{0}}, \abs{q_{1}} \geq \T \right) \wedge (\sign(q_0) = \sign(q_1))\\
1, & \left(\abs{q_{0}}, \abs{q_{1}} \geq \T \right) \wedge (\sign(q_0) \neq \sign(q_1)) \\
?, & \left(\abs{q_{0}} \geq \T, \, \abs{q_{1}} < \T \right) \vee \left(\abs{q_{0}} < \T, \, \abs{q_{1}} \geq \T \right)\\
\varnothing, & \abs{q_{0}}, \abs{q_{1}} < \T
\end{cases},
$$
where ${\T}$ is the threshold value that is fixed before executing the protocol. When $\B \in \{0,1\}$, we say that the outcome is \textit{conclusive}. On the other hand, when $\B = ?$, we say that the outcome is \textit{inconclusive} and when $\B = \varnothing$, we say that there is \textit{no-click}.

\vspace{-0.2cm}
\item \textit{Parameter estimation:} For each round, if $\B = ?$ or $\B = \varnothing$, Bob will reveal $q_\beta$ (for each $\beta$) and ask Alice to reveal her bit value $\A$. Furthermore, if Alice chooses the $X$-basis or $\mu \neq \mu_0$, she would also reveal her bit value $\A$ and ask Bob to reveal his decoded bit $\B$ and his measurement outcome $q_\beta$. After that, Alice would randomly sample a few of the remaining rounds where she reveals her bit value $\A$ and asks Bob to reveals his bit value $\B$ and his measurement outcome $q_\beta$. The remaining rounds are used as their raw keys.

From these announcements, for each intensity setting $\mu_i$ and basis choice $x$, they can estimate the \textit{gain}, i.e., the probability of obtaining conclusive outcome conditioned on Alice's intensity setting and basis choice
\begin{equation*}
    Q_{\mu_i}^x = \Pr[\B \in \{0,1\} | \mu = \mu_i, \X = x],
\end{equation*}
as well as the \textit{quantum bit-error rate} (QBER)
\begin{equation*}
    E_{\mu_i}^x = \Pr[\A \neq \B | \mu = \mu_i, \X = x, \B \in \{0,1\}],
\end{equation*}
and the more fine-grained \textit{behaviour}, i.e., the set of conditional probabilities
\begin{equation*}
    G_{\mu_i}^{(b|a,x)} = \Pr[\B = b| \mu = \mu_i, \A=a, \X = x].
\end{equation*}
They could also estimate the probability of obtaining discretised bin $\nu_\beta = \nu$, conditioned on Alice's basis choice, bit value and intensity setting
\begin{equation*}
    W_{\nu|\mu_i,a,x}^\beta = \Pr[\nu_\beta = \nu|\mu = \mu_i, \A = a, \X = x].
\end{equation*}
Lastly, Bob could estimate the average value of $q_{\beta}^{\, 2}$ conditioned on Alice's basis choice, bit value and intensity settings. They can then calculate
\begin{equation*}
    \omega^{(\beta, a,x)}_i = \frac{\langle q_{\beta}^{\, 2} \rangle_{\mu_i, a, x} -1}{2}.
\end{equation*}

 If these estimated quantities lie within the tolerated intervals that are fixed before executing protocol, then they continue to the next step of the protocol, else, they abort the protocol.

\item \textit{Classical post-processing}: Alice and Bob will employ a \textit{reverse-reconciliation} error-correction protocol as well as privacy amplification to obtain a pair of identical and secret key.
\end{enumerate}

\section{Security analysis}\label{sec: security analysis}
In this section, we analyse the security of our proposed protocol. To simplify the analysis, in this work, we assume that Eve performs \textit{collective attacks} (when she attacks identically and independently in each round) and we restrict our analysis to the \textit{asymptotic} limit (when $N \rightarrow \infty$). We leave the security analysis against the general attacks in the finite-key setting for future work.

Furthermore, throughout this paper, we are working in the \textit{device-dependent} setting. More precisely, we assume that Alice's source will emit the exact states that are specified by the protocol. As such, we have to assume that sufficient isolation and filtering are provided to prevent side-channel attacks such as the Trojan horse attacks or leakages through other degrees-of-freedom. If such side-channels are present, one could adopt the techniques from Refs.~\cite{zhang_powerlimiter_2021, pereira2019quantum, pereira2020quantum} to take those side-channels into account in the security analysis. On Bob's side, in analysing the security of the protocol, we assume that Bob is performing a shot-noise-limited balanced homodyne measurement and any imperfections therein are well characterised. It is also important to ensure that any detector vulnerabilities (such as the one demonstrated in Ref.~\cite{qin_homodyne-detector-blinding_2018}) are appropriately addressed.

Since we have limited our analysis to the asymptotic limit, we could neglect statistical fluctuations in the parameter estimation as well as we can consider the case where $p_Z \rightarrow 1$ and the fraction of rounds used for parameter estimation can be taken to be almost zero. The asymptotic secret key rate $R$ of the protocol under these assumptions are given by the Devetak-Winter bound~\cite{devetak2005distillation}
\begin{multline} \label{eq: devetak-winter}
    R = p_Z p_{\mu_0} Q_{\mu_0}^Z \Big[ H(\B|E, \X = Z, \mu = \mu_0, \B \in \{0,1\}) \\
    - H(\B|\A, \X = Z, \mu = \mu_0, \B \in \{0,1\}) \Big],
\end{multline}
where $Q_{\mu_0}^Z$ is the gain, conditioned on $\mu = \mu_0$ and $\X = Z$ and $H(\cdot)$ denotes the von Neumann entropy. The first term quantifies Eve's uncertainty about Bob's bit values whereas the second term quantifies Alice's uncertainty about Bob's bit value. The latter can be bounded in terms of the QBER
\begin{equation}
    H(\B|\A, \X = Z, \mu = \mu_0, \B \in \{0,1\}) \leq  h_2\left(E_{\mu_0}^Z\right),
\end{equation}
where $h_2(x) = -x \log_2(x) -(1-x) \log_2 (1-x)$ is the binary entropy function.

\subsection{Virtual entanglement-based protocol}
As such, our task is to find a lower bound on the first term of \eqref{eq: devetak-winter}. To that end, we consider a virtual entanglement-based protocol which, from the point-of-view of Eve, is indistinguishable from the actual protocol. In other words, the classical and quantum side-information that Eve has in the actual protocol would be the same as the ones she holds in the virtual protocol.

In the actual protocol, Alice utilises a phase-randomised laser which emits Poissonian mixture of photon number states. For a given intensity setting $\mu$, she emits $n$-photons with probability $p_{n|\mu} = e^{-\mu} \mu^n / n!$. To convert this to an entanglement-based protocol, we could replace this source with the following entangled state which Alice prepares with probability $p_{n|\mu}$
\begin{equation*}
    \ket{\Phi_n}_{\X AA'} = \sqrt{p_Z} \ket{Z}_{\X} \ket{\Phi^Z_n}_{AA'} + \sqrt{1-p_Z} \ket{X}_{\X} \ket{\Phi^X_n}_{AA'},
\end{equation*}
where $\{\ket{Z}_{\X}, \ket{X}_{\X} \}$ are orthogonal states that encode Alice's basis choice and
\begin{equation}
\begin{split}
    \ket{\Phi^Z_n}_{AA'} &= \frac{\ket{0}_A \ket{0_Z}^n_{A'} + \ket{1}_A \ket{1_Z}^n_{A'}}{\sqrt{2}} \\
    \ket{\Phi^X_n}_{AA'} &= \frac{\ket{0}_A \ket{0_X}^n_{A'} + \ket{1}_A \ket{1_X}^n_{A'} }{\sqrt{2}}.
\end{split}
\end{equation}
Here, the states $\{\ket{0}_A, \ket{1}_A\}$ are orthogonal states that encode her bit value. On the other hand, $\{\ket{0_Z}^n_{A'},\ket{1_Z}^n_{A'},\ket{0_X}^n_{A'},\ket{1_X}^n_{A'} \}$ are $n$-photon BB84 states. More precisely, denoting the vacuum state as $\ket{v}$ and defining the orthogonal annihilation operators $\ah_0$ and $\ah_1$ such that $[\ah_0, \ah_1^\dagger] = 0$, for $j \in \{0,1\}$, we have
\begin{equation}
\begin{split}
    \ket{j_Z}^n_{A'} &= \frac{(\ah_j^\dagger)^n}{\sqrt{n!}} \ket{v}, \\
    \ket{j_X}^n_{A'} &= \frac{(\ah_0^\dagger + (-1)^j \ah_1^\dagger)^n}{\sqrt{2^n n!}} \ket{v}.
\end{split}
\end{equation}
Hence, $\ket{0_Z}^n_{A'}$ and $\ket{1_Z}^n_{A'}$ are Fock states in the $\ah_0$ and $\ah_1$ mode, meanwhile $\ket{0_X}^n_{A'}$ and $\ket{1_X}^n_{A'}$ are Fock states in the $\ah_+$ and $\ah_-$ mode, respectively.

Next, Alice sends the quantum system $A'$ to Bob via an untrusted quantum channel and then measures the register $\X$ and $A$ in their corresponding standard basis. This is equivalent to Alice randomly choosing her basis and bit value from the appropriate probability distributions. One could see that by projecting the systems $\X$ and $A$ to the appropriate states, we get the same signal states that we have in the actual prepare-and-measure protocol (for a given photon number $n$).

We can then describe Eve's attack by a map $\E^{(n)}_{A' \rightarrow B}$. In passing, we remark that since Alice sends states that are block diagonal in the photon number basis, Eve could in principle perform a photon number measurement without perturbing the states that Alice sends. Hence, the map may, in general, depend on the emitted photon number $n$. The channel $\E^{(n)}_{A' \rightarrow B}$ would then map the pure state $\ketbra{\Phi_n}{\Phi_n}_{\X AA'}$ to a mixed state $\rho^{(n)}_{\X AB}$. Since Eve is free to subtract/add photons from/to the channel, the number of photons that Bob receive may differ from the one emitted from Alice's source. Thus, the system $B$, in general, lives in an infinite-dimensional Hilbert space. On the other hand, since Alice's virtual systems $\X$ and $A$ are stored securely in her lab, we have the following constraint:
\begin{equation} \label{eq: Alice's state}
    \rho^{(n)}_{\X A} = \Tr_B [\rho^{(n)}_{\X A B}] = \Tr_{A'} \left[ \ketbra{\Phi_n}{\Phi_n}_{\X A A'}\right]
\end{equation}

Finally, upon receiving system $B$, Bob performs homodyne detection on it. The rest of the protocol (Step 3 to Step 5) is identical to the actual protocol described in Section \ref{sec: protocol}.

\subsection{State and measurements: block-diagonal structure}
Now, we turn our attention to Bob's measurement. For simplicity, we shall consider the case in which Bob possesses an ideal homodyne detector. As we shall see in the Appendix \ref{sec: imperfect detector}, the same conclusion could be derived when he uses an imperfect homodyne detector.

Since Bob randomises the global phase of his LO, his measurement would be block-diagonal in the photon number basis. To see this, for a given value of $q_0$ and $q_1$ and LO phase $\varphi$, Bob's POVM element for those particular outcomes is given by
\begin{equation} \label{eq: integral phase random}
    \Pi(q_0,q_1) = \int_{0}^{2\pi} \frac{\diff{\varphi}}{2\pi} \ketbra{q_0(\varphi)}{q_0(\varphi)} \otimes \ketbra{q_1(\varphi)}{q_1(\varphi)},
\end{equation}
where $\ket{q_j(\varphi)}$ is the eigenstate of the quadrature operator
\begin{equation*}
    \hat{Q}_\varphi^{(j)} = \ah_j e^{-i \varphi} + \ah_j^\dagger e^{i \varphi}.
\end{equation*}

We could re-write $\ket{q_j(\varphi)}$ in the photon number basis~\cite{vogel2006quantum}
\begin{equation}
    \ket{q_j(\varphi)} = \sum_{m = 0}^\infty \psi_{m}(q_j) e^{-i m \varphi} \ket{m}, 
\end{equation}
where $\{ \ket{m} \}_m$ are Fock states and
\begin{equation} \label{eq: Fock state wavefunction}
    \psi_m(q_j) = \frac{1}{\sqrt{2^m m! \sqrt{2\pi} }} H_m(q_j/\sqrt{2}) e^{-q_j^2/4}
\end{equation}
is the wavefunction of the Fock state $\ket{m}$ in coordinate representation. Then, we could perform the integration in Eq.~\eqref{eq: integral phase random} to obtain the block-diagonal structure
\begin{multline} \label{eq: block-diagonal measurement}
    \Pi(q_0, q_1) = \sum_{k_0,k_1,l_0,l_1 = 0}^{\infty} \psi_{k_0}(q_0) \psi_{l_0}(q_0) \psi_{k_1}(q_1)\psi_{l_1}(q_1) \\ \delta_{k_0+k_1, l_0 + l_1} \ketbra{k_0,k_1}{l_0, l_1},
\end{multline} 
where we used the following identity
\begin{equation*}
    \int_{0}^{2\pi}\frac{\diff{\varphi}}{2\pi} e^{-i(k_0 + k_1 - l_0 - l_1) \varphi} = \delta_{k_0+k_1, l_0 + l_1}.
\end{equation*}
We can simplify the summation in Eq.~\eqref{eq: block-diagonal measurement}. Making use of the Kronecker delta in the summation, we have
\begin{align}
    &\Pi(q_0, q_1) \nonumber \\
    &= \sum_{m=0}^{\infty} \sum_{k_0 = 0}^{m} \sum_{l_0 = 0}^{m} \psi_{k_0}(q_0) \psi_{l_0}(q_0) \nonumber \\
    & \hspace{2cm}\psi_{m - k_0}(q_1)\psi_{m - l_0}(q_1) \ketbra{k_0, m - k_0}{l_0, m - l_0} \nonumber \\ & =:
    \sum_{m = 0}^\infty \Pi^{(m)}(q_0, q_1).
\end{align}

Now, denoting the total photon number $m = k_0 + k_1 = l_0 + l_1$, due to the block-diagonal structure, we have
\begin{align}
    \Pi(q_0, q_1) &= \sum_{m = 0}^\infty P_m \Pi(q_0, q_1) P_m \nonumber\\
    &=\sum_{m = 0}^\infty P_m \Pi^{(m)}(q_0, q_1) P_m
\end{align}
where $P_m$ is the projection to the $m$-photon subspace (the space in which Bob receives a total of $m$-photons). In other words, we could interpret Bob's measurement as a virtual photon number measurement followed by a reduced measurement $\Pi^{(m)}(q_0, q_1) = P_m \Pi(q_0, q_1) P_m $ which lives in the $m$-photon subspace.

Since Bob's measurement is indistinguishable to the virtual measurement that we have just described, Eve is not penalised if she performs the projection herself. Hence, without loss of generality, we can consider the state shared by Alice and Bob has the following block-diagonal structure
\begin{equation} \label{eq: block-diagonal1}
    \rho^{(n)}_{\X A B} = \bigoplus_{m = 0}^\infty \tilde{\rho}_{\X A B}^{(m,n)} 
\end{equation}
where $\tilde{\rho}_{\X A B}^{(m,n)}$ is a (sub-normalised) state with Bob receiving $m$ photons distributed across the two temporal modes when Alice sends $n$ photons. If the normalised version of $\tilde{\rho}_{\X A B}^{(m,n)}$ is denoted by $\rho_{\X A B}^{(m,n)}$, we have
\begin{equation}
    \tilde{\rho}_{\X A B}^{(m,n)} = q_{m|n} \rho_{\X A B}^{(m,n)},
\end{equation}
with a normalisation factor $q_{m|n}$. The normalisation factor $q_{m|n}$ can therefore be interpreted as the probability of Bob receiving $m$-photons, conditioned on Alice sending $n$-photons. Importantly, we can find upper and lower bounds on $q_{m|n}$ using a variant of decoy-state method proposed in Ref.~\cite{lavie2021estimating} which we will discuss in Section~\ref{sec: decoy}.

\subsection{Refined Pinsker's inequality} \label{sec: pinsker}
Now, to bound Eve's uncertainty about Bob's measurement results, we want to find a lower bound to the conditional von Neumann entropy $H(\B|E, \X = Z, \mu = \mu_0, \B \in \{0,1\})$. We have
\begin{multline*}
    H(\B|E, \X = Z, \mu = \mu_0, \B \in \{0,1\})  \\
    \geq H(\B|E,\X = Z, \mu = \mu_0, \B \in \{0,1\}, \N, \M),
\end{multline*}
where $\N$ and $\M$ denotes the input and output photon number, respectively. We could further lower bound the above by
\begin{multline*}
Q_{\mu_0}^Z H(\B|E,\X = Z, \mu = \mu_0, \B \in \{0,1\}, \N, \M) \\
\geq  \sum_n p_{n|\mu_0} q_{m \leq n | n} s_{m \leq n|n}
H(\B|E, \N = n, \postselect),
\end{multline*}
where $q_{m\leq n|n} = \Pr[\M \leq n| \N = n]$, $s_{m \leq n|n} = \Pr[\B \in \{0,1\}|\M \leq n, \N = n]$ and for brevity, we call the event $(\X = Z, \mu = \mu_0, \B \in \{0,1\}, \M \leq \N)$ as `postselected' (in short, $\postselect$).

\correction{In passing, we remark that the honest parties do \textit{not} postselect on rounds in which $\M \leq \N$ (i.e., rounds in which the number of photons that Bob receives is not more than the one prepared by Alice). In the protocol, neither Alice nor Bob know the number of photons that they prepare or receive. The only postselection that is being performed in the protocol is with respect to whether Bob's measurement outcome is conclusive. However, one can conservatively extract secrecy from rounds in which $\M \leq \N$ and compute the conditional entropy based on this event. The effect is identical to a hypothetical scenario where Alice and Bob perform non-demolition photon number measurement on their optical systems and postselecting on the event in which $\M \leq \N$.}

Following the argument from Ref.~\cite{schwonnek2021device}, we can lower bound $H(\B|E, \N = n, \postselect)$ as
\begin{equation}
    H(\B|E, \N = n, \postselect) \geq p_{\postselect}^{(n)} \left[ 1 - h_2 \left(\frac{1 - V_n}{2} \right)\right],
\end{equation}
where $p_{\postselect}^{(n)} = \Pr[\B \in \{0,1\} , \M \leq \N| \N = n, \X = Z, \mu = \mu_0]$ and
\begin{equation} \label{eq: trace norm}
    V_{n} = \norm{\sigma^{(n)}_{A B R} - \pinching \left[ \sigma^{(n)}_{A B R}  \right]}_1,
\end{equation}
with $\pinching$ is the pinching channel associated to Bob's measurement, and $\sigma^{(n)}_{ABR}$ is the Naimark's dilated state, when postselected on the conclusive event $\B \in \{0,1\}$ and $\M \leq n$. The explicit form of the state $\sigma^{(n)}_{ABR}$ is derived in Appendix~\ref{sec: postselection and pinching}.

Hence, our task is to find a lower bound on the trace-norm $V_n$. This can be formulated as a semidefinite program (SDP). Denoting $\Lambda := \sigma^{(n)}_{A B R} - \pinching \left[ \sigma^{(n)}_{A B R}\right]$, we have
\begin{equation} \label{eq: sdp1}
\begin{split}
    V_{n} = \min_{Y_1, Y_2, \rho_{\X A B}^{(n)}} \ &\frac{1}{2} \Tr[Y_1 + Y_2]\\
        \text{s.t.} \quad 
        &\begin{pmatrix}
        Y_1 & \Lambda\\
        \Lambda^\dagger & Y_2
        \end{pmatrix} \succeq 0,\\
        & \Tr[\rho^{(n)}_{\X A B}] = 1,\\
        & \Tr_{B}[\rho^{(n)}_{\X A B}] = \rho^{(n)}_{\X A},\\
        & \rho^{(n)}_{\X A B} \in \Sset_n,
\end{split}
\end{equation}
where $\Sset_n$ is the set of density matrices that could reproduce the statistics observed in the parameter estimation step. The set $\Sset_n$ can be characterised by functions that are linear in $\rho_{\X A B}^{(n)}$. We have
\begin{equation} \label{eq: equality constraints}
\begin{split}
    \Tr[ \ketbra{x}{x} \otimes \ketbra{a}{a} \otimes M_{b|x}^{(m \leq n)} \, \rho^{(n)}_{\X A B}] &= \frac{p_x}{2} \Gamma_{m \leq n}^{(b|a,x)},\\
    \Tr[ \ketbra{x}{x} \otimes \ketbra{a}{a} \otimes P_{m} \, \rho^{(n)}_{\X A B}] &= \frac{p_x}{2} q^{(a,x)}_{m|n},
\end{split}
\end{equation}
for all values of $(a,b,x,n)$ and for all $m \leq n$, where $\Gamma_{m \leq n}^{(b|a,x)} = \Pr[\B = b, \M \leq n| \A = a, \X = x, \N = n]$, $q_{m|n}^{(a,x)} = \Pr[\M = m| \A = a, \X = x, \N=n]$  and $M_{b|x}^{(m \leq n)}$ is Bob's measurement operators (for the space $\M \leq n$) obtained by integrating $\sum_{m \leq n} \Pi^{(m)}(q_0,q_1)$ over the appropriate intervals defined by the decoding scheme mentioned in Section \ref{sec: protocol}. The operator $P_{m}$ denotes the projector on the $m$-photon subspace of Bob's system.

The constraints \eqref{eq: equality constraints} is a characterisation of the set $\Sset_n$. However, $\Gamma_{m \leq n}^{(b|a,x)}$ and $q_{m|n}$ are not directly observed in the experiment. Fortunately, it is possible to establish upper and lower bounds on them by using a variant of the decoy-state method~\cite{lo_decoy_2005, lavie2021estimating}
(Section~\ref{sec: decoy}). For now, it suffices to assume that one can obtain bounds of the type
\begin{equation} \label{eq: inequality constraints}
\begin{split}
    \Gamma_{m \leq n}^{(b|a,x),L} \leq \Gamma_{m \leq n}^{(b|a,x)} &\leq \Gamma_{m \leq n}^{(b|a,x),U}, \\
    q^{(a,x),L}_{m|n} \leq \hphantom{\gamma} q^{(a,x)}_{m|n} \hphantom{\gamma} &\leq q^{(a,x),U}_{m|n}.
\end{split}
\end{equation}

By replacing the equality constraints \eqref{eq: equality constraints} by the inequality constraints \eqref{eq: inequality constraints}, we construct a set $\Sset_n'$ such that $\Sset_n \subset \Sset_n'$. Hence, this provides a relaxation for the SDP \eqref{eq: sdp1} that would yield a valid lower bound on $V_n$. 

Unfortunately, even the relaxed SDP is still computationally intractable due to Bob's system being infinite dimensional.
However, we shall prove that without loss of generality, one could consider a state with finite rank and hence reduces the infinite-dimensional SDP into a finite-dimensional one.

To that end, suppose that $\rho_{\text{opt}}$ is an optimiser of the infinite-dimensional SDP. As mentioned previously, without loss of generality, we can consider states that are block-diagonal in the photon number basis. As such, we can write $\rho_{\text{opt}}$ as
\begin{equation}
    \rho_{\text{opt}} = q_{m\leq n |n} \, \rho^{(m \leq n)}_{\text{opt}} \, \oplus\,  q_{m > n |n} \, \rho^{(m > n)}_{\text{opt}},
\end{equation}
where the state $\rho^{(m \leq n)}_{\text{opt}}$ lives in the subspace where Bob's photon number is $m \leq n$ whereas the state $\rho^{(m > n)}_{\text{opt}}$ lives in the $m > n$ subspace. Then, $q_{m \leq n |n}$ gives the probability of obtaining $m \leq n$ whereas  $q_{m > n|n}$ gives the probability of obtaining $m > n$, both conditioned on Alice sending $n$-photons. Clearly, $\rho^{(m \leq n)}_{\text{opt}}$ is finite-dimensional.

Now, consider the map $\ps$ that corresponds to the postselection in the protocol, i.e., $\ps[\rho^{(n)}_{\X A B}] = \sigma_{A B R}^{(n)}$. Since we only consider secrecy when $\M \leq \N$, if Alice emits $n$ photons, we have $\ps[\rho_{m > n}] = 0$ for any state $\rho_{m > n}$ that lives in the $m > n$ subspace. Then, consider the state
\begin{equation}
    \rho'_{\text{opt}} = q_{m\leq n |n} \, \rho^{(m \leq n)}_{\text{opt}} \, \oplus\,  q_{m > n |n}.
\end{equation}
By construction, if $\rho_\text{opt} \in \Sset'_n$, we also have 
$\rho'_\text{opt} \in \Sset'_n$ since the constraints only depend on the part of the states where $m \leq n$. Secondly, we have $\ps[\rho_\text{opt}] = \ps[\rho'_\text{opt}]$, which implies that the two states share the same optimal value of $V_n$. This implies that there exists a finite-dimensional optimiser for the infinite-dimensional SDP. Hence, without loss of generality, we can consider a finite-dimensional version of the SDP \eqref{eq: sdp1}. In particular, if the $m \leq n$ subspace has a dimension of $\dim(\Hset_{m \leq n}) = \sum_{m = 0}^n (m+1)$, it is sufficient to consider the dimension of Bob's system to be $d_B = \dim(\Hset_{m \leq n}) + 1$. Hence, taking into account the relaxation and the dimension reduction, we have the following SDP:
\begin{equation} \label{eq: sdp2}
\begin{split}
    V_{n} = \min_{Y_1, Y_2, \rho_{\X A B}^{(n)}} \ &\frac{1}{2} \Tr[Y_1 + Y_2]\\
        \text{s.t.} \quad 
        &\begin{pmatrix}
        Y_1 & \Lambda\\
        \Lambda^\dagger & Y_2
        \end{pmatrix} \succeq 0,\\
        & \Tr[\rho^{(n)}_{\X A B}] = 1,\\
        & \Tr_{B}[\rho^{(n)}_{\X A B}] = \rho^{(n)}_{\X A},\\
        & \rho^{(n)}_{\X A B} \in \Sset'_n, \\
        & \dim\left( \rho^{(n)}_{B}  \right) = \dim(\Hset_{m \leq n}) + 1.
\end{split}
\end{equation}

\subsection{Estimating the channel behaviour and its input-output photon number distribution}\label{sec: decoy}
To formulate the SDP \eqref{eq: sdp2}, we need a characterisation of the set $\Sset'_n$ as linear functions of the state $\rho_{\X A B}^{(n)}$. As discussed in the previous section, this can be done if we have lower and upper bounds as written in Eq.~\eqref{eq: inequality constraints}. We will find these bounds using the decoy-state method \cite{lo_decoy_2005}.

Recall that Alice prepares phase-randomised coherent states which are Poissonian mixtures of Fock states. Assuming that the phase-randomisation is done properly, given a Fock state, Eve would not be able to deduce the intensity setting that Alice chose. As such, Eve's attack must not depend on Alice's intensity setting $\mu$. Hence, by using different intensity settings and observing how the resulting statistics depend on the intensity settings, Alice and Bob could estimate the channel behaviour and its input-output photon number distribution.

Now, consider the behaviour, $G_{\mu_i}^{(b|a,x)}$, that Alice and Bob estimate during the protocol. We can expand the behaviour in terms of the $n$-photon behaviours
\begin{equation}
    G_{\mu_i}^{(b|a,x)} = \sum_{n = 0}^{\infty} p_{n|\mu_i} \Gamma_{n}^{(b|a,x)},
\end{equation}
where $\Gamma_{n}^{(b|a,x)} = \Pr[\B = b| \A = a, \X = x, \N = n]$. One could obtain an upper and lower bound on $\Gamma_n^{(b|a,x)}$ via the standard decoy-state method~\cite{lo_decoy_2005}. For example, if we are interested in $\N = n'$, one can formulate the following linear program (LP)
\begin{equation}
    \begin{split}
      \underset{\{\Gamma_n^{(b|a,x)}\}_n}{\text{max/min}} \quad & \Gamma_{n'}^{(b|a,x)}\\
      \text{s.t.} \quad & \sum_{n} p_{n|\mu_i} \Gamma_{n}^{(b|a,x)} = G_{\mu_i}^{(b|a,x)} \qquad \forall \mu_i, \\
      & 0 \leq \Gamma_n^{(b|a,x)} \leq 1 \qquad \forall n.
    \end{split}
\end{equation}
 The above LP involves infinitely many variables $\{\Gamma_n^{(b|a,x)}\}_n$. However, using the fact that $0 \leq \Gamma_n^{(b|a,x)} \leq 1$ for all $n > \nmax$, we can consider a photon number cutoff $\nmax$ and relax the problem
\begin{equation} \label{eq: decoy state LP}
    \begin{split}
      \underset{\{\Gamma_n^{(b|a,x)}\}_n}{\text{max/min}} \quad & \Gamma_{n'}^{(b|a,x)}\\
      \text{s.t.} \quad & \sum_{n=0}^{\nmax} p_{n|\mu_i} \Gamma_{n}^{(b|a,x)} \leq G_{\mu_i}^{(b|a,x)} \qquad \forall \mu_i\\
      &  \sum_{n=0}^{\nmax} p_{n|\mu_i} \Gamma_{n}^{(b|a,x)} \geq G_{\mu_i}^{(b|a,x)} \\
      & \hspace{3cm}- \left(1 - \sum_{n = 0}^{\nmax} p_{n|\mu_i} \right) \quad \forall \mu_i\\
       &0 \leq \Gamma_n^{(b|a,x)} \leq 1 \qquad \forall n \in \{0,...,\nmax\}.
    \end{split}
\end{equation}
Note that, in our relaxation, we have enlarged the feasible region of the original LP such that we obtain a valid lower and upper bounds on $\Gamma_{n'}^{(b|a,x)}$.

In turn, we can also use the law of total probability to expand $\Gamma_{n}^{(b|a,x)}$ and obtain
\begin{equation}
    \Gamma_{n}^{(b|a,x)} = \sum_{m = 0}^\infty q_{m|n}^{(a,x)} \gamma_{m,n}^{(b|a,x)} ,
\end{equation}
where $\gamma_{m,n}^{(b|a,x)} = \Pr[\B = b| \N = n, \M = m, \A = a, \X = x]$. Then, we have
\begin{align}
    \Gamma_{m\leq n}^{(b|a,x)} &= \sum_{m \leq n} q_{m|n}^{(a,x)} \gamma_{m,n}^{(b|a,x)} \nonumber\\
    &= \Gamma_n^{(b|a,x)} - \sum_{m > n} q_{m|n}^{(a,x)} \gamma_{m,n}^{(b|a,x)}
\end{align}
Using $0 \leq \gamma_{m,n}^{(b|a,x)} \leq 1$, we have an upper and lower bound on $\Gamma_{m \leq n}^{(b|a,x)}$
\begin{align}
    \Gamma_{m \leq n}^{(b|a,x)} &\leq \Gamma_n^{(b|a,x)} \label{eq: Gamma upper bound}\\
    \Gamma_{m \leq n}^{(b|a,x)} &\geq \Gamma_n^{(b|a,x)} - \left(1 - q_{m \leq n | n}^{(a,x)} \right) \label{eq: Gamma lower bound}
\end{align}
where $q_{m \leq n|n}^{(a,x)} = \sum_{m \leq n} q_{m|n}^{(a,x)}$.

Therefore, our remaining task is to find upper and lower bounds on the conditional probabilities $q_{m|n}^{(a,x)}$. Here, we use the linear programming method proposed in Ref.~\cite{lavie2021estimating}.

Suppose that Bob receives $m$-photons in mode $\ah_\beta$, if he measures the quadrature $\qh_\varphi^{(\beta)} = \ah_\beta e^{-i \varphi} + \ah_\beta^\dagger e^{i \varphi}$ using a LO with randomised phase, the probability of obtaining an outcome $\nu$ that lies within the interval $I_\nu$ is given by
\begin{equation}
    C_{\nu | m}^\beta = \int_{\nu \delta}^{(\nu+1)\delta} \diff q \, \abs{\psi_m(q)}^2 .
\end{equation}
Let $\ket{m_\beta}$ be the state with photon number $m$, all in the mode $\ah_\beta$, i.e.,
\begin{equation*}
    \ket{m_\beta} = \frac{\left(\ah_{\beta}^\dagger\right)^m}{\sqrt{n!}}  \ket{v}.
\end{equation*}
Then, we have $\ketbra{m_\beta}{m_\beta} \preceq \1_m$. As such, we have
\begin{equation}
    q_{m|n}^{(\beta, a, x)} \leq q_{m|n}^{(a,x)},
\end{equation}
where $q_{m|n}^{(\beta, a, x)} = \Pr[\M = m, \text{mode} = \ah_\beta | \N = n, \A = a, \X = x]$. For upper bounds, we can trivially use $q_{m|n}^{(a,x)} \leq 1$.

Using the chain rule and the law of total probability, we can then write
\begin{equation}
    W^\beta_{\nu|\mu_i, a, x} = \sum_{n=0}^\infty \sum_{m = 0}^\infty p_{n|\mu_i} q_{m|n}^{(\beta, a, x)} C_{\nu|m}^\beta.
\end{equation}
One could also consider the square of the quadrature operator
\begin{equation}
    \left(\qh_\varphi^{(\beta)} \right)^2 = \ah_\beta^{\, 2} e^{-2i \varphi} + \left(\ah_\beta^\dagger\right)^{\, 2} e^{2i \varphi} + \ah_\beta^\dagger \ah_\beta +  \ah_\beta \ah_\beta^\dagger.
\end{equation}
Then randomising the phase, we get
\begin{equation}
    \int_{0}^{2\pi} \frac{\diff \varphi}{2\pi} \, \left(\qh_\varphi^{(\beta)} \right)^2 = \ah_\beta^\dagger \ah_\beta +  \ah_\beta \ah_\beta^\dagger = 2 \ah_\beta^\dagger \ah_\beta + \1.
\end{equation}
Since $\ah_\beta^\dagger \ah_\beta$ is the number operator for the mode $\ah_\beta$, then
\begin{equation*}
    \omega^{(\beta, a,x)}_i = \frac{\langle q_{\beta}^{\, 2} \rangle_{\mu_i, a, x} -1}{2}
\end{equation*}
is the mean photon number in mode $\ah_\beta$ that Bob receives. Suppose we are interested in bounding $q_{m|n}^{(\beta, a, x)}$ for $\M = m'$ and $\N = n'$. Then we can consider the following LP:

\begin{equation}
    \begin{split}
        \underset{{\{q_{m|n}^{(\beta,a,x)}\}_{m,n}}}{\text{min} }\quad & q_{m'|n'}^{(\beta,a,x)} \\
        \text{s.t.}  \quad & \sum_{n,m} p_{n|\mu_i} q_{m|n}^{(\beta,a,x)} C^\beta_{\nu|m} = W^\beta_{\nu|\mu_i,a,x}, \quad \forall \mu_i,\nu\\
        & \sum_{n,m} p_{n|\mu_i} \  q_{m|n}^{(\beta,a,x)} m = \omega_i^{(\beta, a, x)}, \quad \forall \mu_i\\
        & 0 \leq q_{m|n}^{(\beta,a,x)} \leq 1, \quad \forall m,n\\
        & \sum_{m} q_{m|n}^{(\beta,a,x)} = 1, \quad \forall n.
    \end{split}
\end{equation}
Again, this LP involves infinitely many variables which make it intractable. To get a valid relaxation, we choose some cutoffs $\nmax$ and $\mmax$ (for $n$ and $m$, respectively). Then, we use the fact that $0 \leq q_{m|n}^{(\beta, a, x)} \leq 1$ to relax the LP to

\begin{widetext}
\begin{equation} \label{eq: relaxed LP double decoy}
    \begin{split}
        \underset{{\{q_{m|n}^{(\beta,a,x)}\}_{m,n}}}{\text{min} }\quad & q_{m'|n'}^{(\beta,a,x)} \\
        \text{s.t.}  \quad & \sum_{n= 0}^{\nmax} \sum_{m = 0}^{\mmax} p_{n|\mu_i} q_{m|n}^{(\beta,a,x)} C^\beta_{\nu|m} \leq W^\beta_{\nu|\mu_i,a,x}, \quad \forall \mu_i,\nu\\
        & \sum_{n= 0}^{\nmax} \sum_{m = 0}^{\mmax} p_{n|\mu_i} \  q_{m|n}^{(\beta,a,x)} (\mmax + 1 - m) \geq (\mmax+1) \left(\sum_{n= 0}^{\nmax} p_{n|\mu_i} \right) -\omega_i^{(\beta, a, x)}, \quad \forall \mu_i\\
        & 0 \leq q_{m|n}^{(\beta,a,x)} \leq 1, \quad \forall m \in \{0, ..., \mmax\} ,n \in \{0, ..., \nmax\}\\
        & \sum_{m=0}^{\mmax} q_{m|n}^{(\beta,a,x)} \leq 1, \quad \forall n \in \{0, ..., \nmax\},
    \end{split}
\end{equation}
\end{widetext}

 In practice, we only impose the first constraint for $\nu \in \{\nu_\text{min}, ..., \nu_{\text{max}}\}$ for some cutoffs $\nu_{\text{min}}$ and $\nu_{\text{max}}$. Since the relaxed constraints are necessary conditions that the original feasible points must satisfy, the bounds that we obtain from the relaxed LP \eqref{eq: relaxed LP double decoy} are valid lower bounds on $q_{m|n}^{(\beta, a, x)}$. One could also find a lower bound on $q_{m \leq n |n}^{(a,x)}$ by changing the objective function in \eqref{eq: relaxed LP double decoy} to $\sum_{m = 0}^{n} q_{m|n}^{(\beta, a, x)}$.

To summarise, we could find a characterisation to the set of feasible density matrices $\Sset_n'$ in terms of linear functions of the state $\rho_{\X A B}^{(n)}$. To this end, we need to find bounds on $\Gamma_{m \leq n}^{(b|a,x)}$ and $q_{m|n}^{(a,x)}$. To bound the latter, we use the fact that $q_{m|n}^{(\beta, a,x)} \leq q_{m|n}^{(a,x)} \leq 1$. We can, in turn, find a lower bound on $q_{m|n}^{(\beta, a,x)}$ using the LP \eqref{eq: relaxed LP double decoy}. Using a similar LP, we can also bound $q_{m \leq n |n}^{(a,x)}$. On the other hand, to bound $\Gamma_{m \leq n}^{(b|a,x)}$, we need to first find bounds on $\Gamma_n^{(b|a,x)}$. To do that, we can use the standard decoy-state method. This can be done by formulating a LP as shown in \eqref{eq: decoy state LP}. Then, we can plug in these bounds, together with the lower bound on $q_{m \leq n |n}^{(a,x)}$, to Eqs.~\eqref{eq: Gamma upper bound} and \eqref{eq: Gamma lower bound}. Finally, since we have characterised $\Sset'_n$ as constraints that are linear in the state $\rho_{\X A B}^{(n)}$, we can efficiently solve the SDP \eqref{eq: sdp2} using standard SDP solvers \footnote{For example, the results we obtained in Section~\ref{sec: simulation} are obtained by solving the SDP and LP using the solver MOSEK~\cite{mosek} via CVX~\cite{cvx}, a package for specifying convex optimisation problems.}.

\section{Simulation}\label{sec: simulation}
To simulate the performance of the protocol, we assume that the loss in the channel can be modelled by a beam-splitter of transmittivity $\eta$ with
\begin{equation}
    \eta = \eta_\text{det} 10^{- \xi L_{AB} / 10}
\end{equation}
where $\eta_\text{det}$ is the effective efficiency of the homodyne detector, $\xi$ is the fiber loss coefficient in dB/km (for standard fiber, $\xi = 0.2$ dB/km for telecom wavelength) and $L_{AB}$ is the distance between Alice's and Bob's lab.

The probability density function (PDF) for homodyne measurement of a (single-mode) coherent state $\ket{\alpha}$ with local oscillator's phase $\varphi$ is given by
\begin{equation*}
    f(q|\alpha, \varphi) = \frac{1}{\sqrt{2\pi}} \exp\left[\frac{-(q - 2 \abs{\alpha} \cos(\theta - \varphi))^2}{2} \right]
\end{equation*}
where $\theta = \arg(\alpha)$ is the phase of the coherent state. Therefore, given the multimode coherent state $\ket{\alpha_0}\ket{\alpha_1}$, we have
\begin{multline}
        f(q_0,q_1 |\alpha_0, \alpha_1, \varphi) \\
        = \frac{1}{2\pi}
        \exp\left[\frac{-(q_0 - 2 \abs{\alpha_0} \cos(\theta_0 - \varphi))^2}{2} \right] \\
        \exp\left[\frac{-(q_1 - 2 \abs{\alpha_1} \cos(\theta_1 - \varphi))^2}{2} \right]
\end{multline}

To simulate the statistics $G^{(b|a,x)}_{\mu_i}$, we first notice that the channel maps $\abs{\alpha_k} \rightarrow \abs{\alpha_k} \sqrt{\eta}$. Furthermore, we let $\theta_0 = \theta_\text{global}$ and $\theta_1^{(a,x)} = \theta_\text{global} + \theta_{\text{rel}}^{(a,x)}$ since the phase of the early temporal mode is always randomised. The magnitude $\abs{\alpha_0}$ and $\abs{\alpha_1}$ are determined by the intensity setting $\mu$ as well as Alice's random inputs $(a,x)$. We have
\begin{multline}
    G^{(b|a,x)}_{\mu_i} = \int_{I_{(b,x)}} \diff{q_0} \diff{q_1} \\
    \int_{0}^{2\pi} \frac{\diff{\varphi}}{2\pi} \ f(q_0,q_1 |\alpha_0^{(a,x,\mu)} \sqrt{\eta}, \alpha_1^{(a,x,\mu)} \sqrt{\eta}, \varphi)
\end{multline}
where the integration over $q_0$ and $q_1$ is carried out over intervals that depend on $b$ and $x$ as determined by the decoding functions. Since both the global phase $\theta_\text{global}$ and the local oscillator phase are randomised and the PDF only depends on their difference, we would observe the same statistics if we fix the global phase $\theta_\text{global}$ and only randomise the phase of the local oscillator $\varphi$.

Similarly, we can calculate $W^\beta_{\nu|\mu_i,a,x}$ using
\begin{equation}
    W^\beta_{\nu|\mu_i,a,x} = \int_{\nu \delta}^{(\nu+1)\delta} \diff{q}  \int_{0}^{2\pi} \frac{\diff{\varphi}}{2\pi} \ f(q| K_{\beta,a,x} \sqrt{\mu_i \eta}, \varphi),
\end{equation}
where we have $[\ah_{(a,x)}, \ah_{\beta}^\dagger] = K_{\beta, a, x} \1$ and $\ah_{(a,x)}$ is the mode associated to $\A = a$ and $\X = x$ and $\ah_\beta$ is the mode associated to $\beta$.

\begin{figure}[h]
    \centering
    \includegraphics[width = \columnwidth]{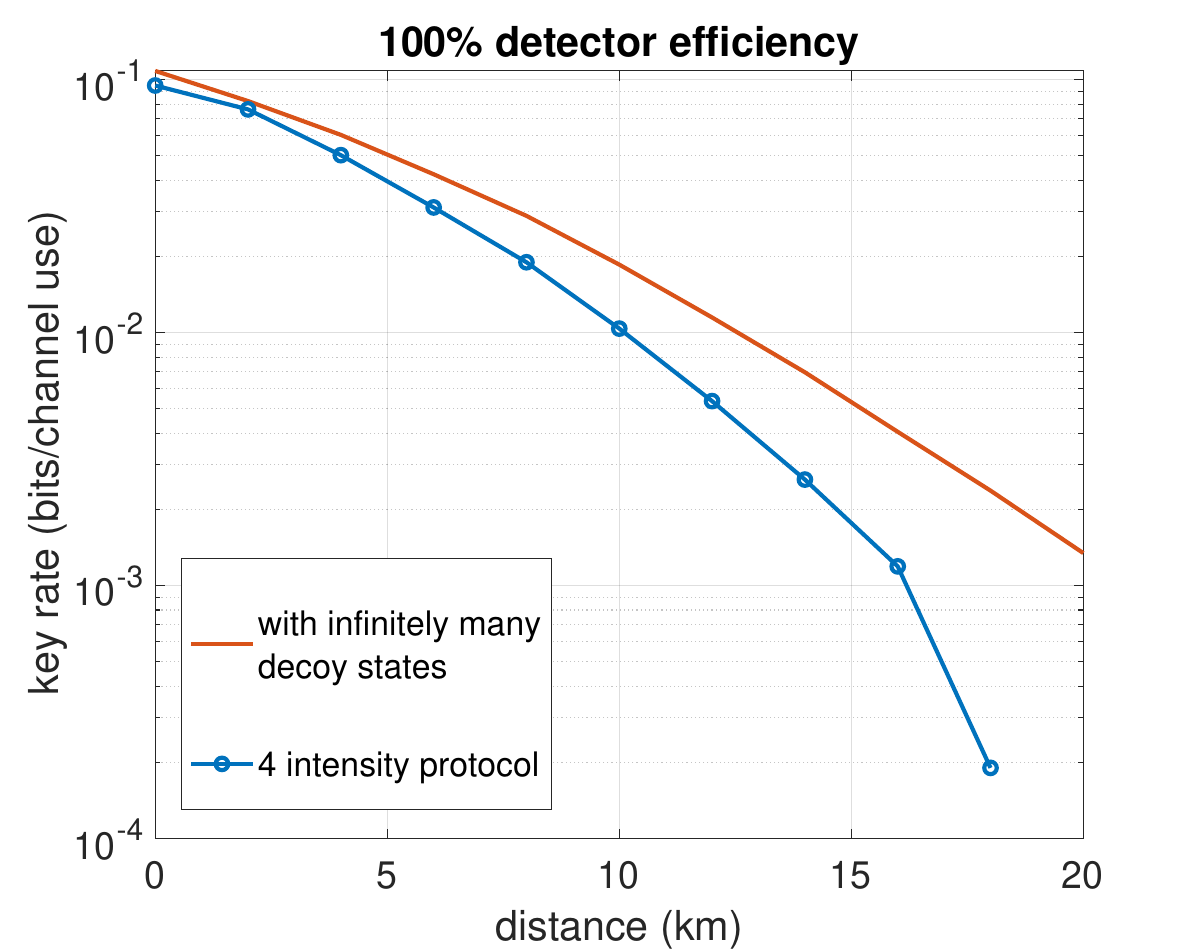}
    \caption{\textbf{Key rate vs distance between Alice and Bob for 100\% detector efficiency}. We compare the performance of our proposed four intensity protocol against the case in which we use infinitely many decoy states (such that Alice and Bob could characterise $q^{(a,x)}_{m|n}$ and $\Gamma^{(b|a,x)}_{m \leq n}$ exactly instead of bounding them).}
    \label{fig: key rate plot 100}
\end{figure}

Using this model, we first plot the secret key rate against the distance between Alice and Bob assuming a perfect detection efficiency $\eta_\mathrm{det}$. In our simulation, we consider the case where Alice and Bob only extract secrecy from $n = 0, 1, 2$. The result is plotted in Fig.~\ref{fig: key rate plot 100}.

\begin{table*}[t]
    \centering
    \begin{tabular}{c c c c c c}
    \toprule
         distance & \multicolumn{5}{c}{parameters} \\
         \cmidrule{2-6}
         (km) & $\mu_0$ & $\mu_1$ & $\mu_2$ & $\mu_3$  & $\tau$ \\
     \hline
         0 & 1.200 & $2.3 \times 10^{-3}$ & $5.0 \times 10^{-4}$ & $1.0 \times 10^{-4}$  & 1.600\\
         5 & 0.974 & $2.6 \times 10^{-3}$ & $5.2 \times 10^{-4}$ & $1.0 \times 10^{-4}$  & 1.999\\
         10 & 0.568 & $1.8 \times 10^{-3}$ & $2.7 \times 10^{-4}$ & $1.0 \times 10^{-4}$  & 2.193\\
         15 & 0.347 & $2.1 \times 10^{-3}$ & $3.0 \times 10^{-4}$ & $1.0 \times 10^{-4}$  & 3.314\\
     \bottomrule
    \end{tabular}
    \caption{\textbf{Protocol parameters used in the trusted detector efficiency scenario in Fig.~\ref{fig: key rate plot 72}}. Here, we fix the lowest intensity $\mu_3$ and heuristically optimise the other parameters using a grid search optimisation.}
    \label{tab: parameters}
\end{table*}

 As one can see, in the short distance regime ($\lesssim 5$~km), the four-intensity protocol is almost optimal. However, as the distance between Alice and Bob increases, the gap between the four-intensity protocol and the infinite decoy-state protocol increases as well. From our numerical investigation, this is mainly caused by the two-photon decoy-state bounds which tend to be loose as the loss increases. One possible way to circumvent this is to increase the number of decoy states.

\begin{figure}[h]
    \centering
    \includegraphics[width = \columnwidth]{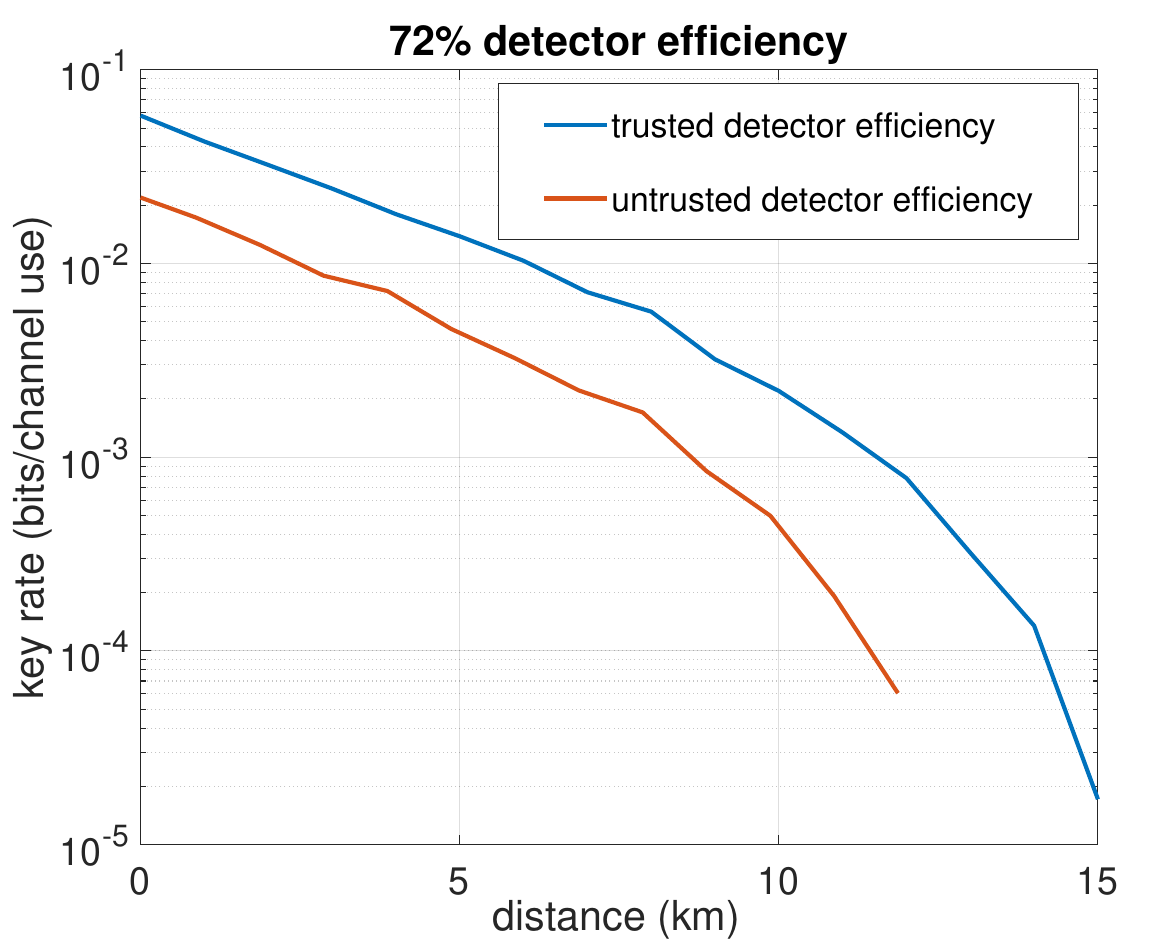}
    \caption{\textbf{Key rate vs distance between Alice and Bob for 72\% detector efficiency}. We compare the performance of our proposed four-intensity protocol under the trusted detector efficiency scenario against the scenario in which the detector loss is attributed to Eve. 
    }
    \label{fig: key rate plot 72}
\end{figure}

To assess the practicality of the protocol, we also perform a simulation for the case when Bob does not use a perfect homodyne measurement. Such imperfections may include imperfect quantum efficiency and electronic noise. However, following the argument of Ref.~\cite{appel2007electronic}, one could see an independent Gaussian electronic noise as an additional loss. As such, in our simulation, we assume that Bob has a lossy homodyne detector with an effective efficiency of the homodyne detector $\edet = 72\%$.
The result is shown in Fig.~\ref{fig: key rate plot 72}.

As one can see from Fig.~\ref{fig: key rate plot 72}, by trusting the detector imperfection, we can increase the key rate for a given distance and improve the robustness of the protocol against loss. However, the protocol could not tolerate high loss even when the effective detector efficiency is trusted.

\correction{To illustrate the range of parameters that are used in the simulation of Fig.~\ref{fig: key rate plot 72} with trusted detector efficiency, we present the heuristically optimised protocol parameters for selected transmission distances in Table.~\ref{tab: parameters}}

\section{Discussion} \label{sec: discussion}
As we can see from Section~\ref{sec: simulation}, our protocol is suitable for high-speed QKD across metropolitan distances. This feature is shared with the protocol proposed in Ref.~\cite{qi_bb84_2021}. Furthermore, some aspects of the security analysis presented in this work could also be adapted to Ref.~\cite{qi_bb84_2021}'s protocol with some minor adjustments. For example, one could also apply the SDP formulation from the refined Pinsker's inequality presented in Section~\ref{sec: pinsker} and the channel estimation technique described in Section~\ref{sec: decoy} to analyse the security of Ref.~\cite{qi_bb84_2021}'s protocol. Nevertheless, the dimension reduction used in this work relies heavily on the block-diagonal structure of the measurement that Bob performs. As such, to prove the security of Ref.~\cite{qi_bb84_2021}'s protocol, one might need to use other methods of reducing Bob's dimension. For example, one can use the technique presented recently in Ref.~\cite{upadhyaya_dimension_2021}.

Next, we also remark that while our security analysis involved some dimension reduction on Bob's system, this dimension reduction is done by deriving necessary conditions that any feasible solution to the original infinite-dimensional optimisation problem must satisfy. This results in a relaxation where the feasible region of the relaxed optimisation problem contains the feasible region of the original optimisation problem. Furthermore, due to the block-diagonal structure of Bob's measurement, we could show that without loss of generality, it is sufficient to consider finite-dimensional states. This is not the case for the security analyses presented in Ref.~\cite{lin2019asymptotic, ghorai2019asymptotic}. In those cases, one could obtain a valid lower bound on the secret key rate by further adopting the technique of Ref.~\cite{upadhyaya_dimension_2021}, but at a price of some correction term due to the cutoffs.

\correction{
It is also remarkable that our dimension reduction bares some resemblance with the technique presented in Ref.~\cite{zhang2021security}. Similar to what we have done here, the authors of Ref.~\cite{zhang2021security} leveraged on the block-diagonal structure of the measurement and argue that one can reduce the analysis to finite dimensional convex optimisation problem. To summarise the argument, they formulated the so-called flag-state squasher which is a quantum channel that maps the infinite-dimensional quantum state to a finite-dimensional one. Then, one can modify the measurement operators such that they only span the finite-dimensional subspace in which the squashed state lives. Furthermore, for any state with the block-diagonal structure in the photon number basis, applying the modified measurement on the squashed state is equivalent to applying the original measurement to the original state.}

\correction{
While our method also exploits the fact that the measurements are block-diagonal in the photon number basis, our argument to reduce the dimension of the optimisation problem is different from the method presented in Ref.~\cite{zhang2021security}. To be precise, the constraints that we use in our optimisation are based on the bounds on the statistics for the case in which Bob does not receive more photons than the ones prepared by Alice. Then, we argue that there exists a finite-dimensional state that is an optimal solution to our optimisation problem. In contrast, Ref.~\cite{zhang2021security} formulated the constraints based on the statistics of the full infinite-dimensional measurements and hence, they have to construct the flag-state squasher to account for the contribution of the higher photon number subspace.}

Finally, while we propose a protocol with reverse reconciliation in this work, one could also consider performing direct reconciliation. However, our initial findings show that the variant with direct reconciliation has less robustness against loss compared to our proposed protocol. However, interestingly, the direct reconciliation protocol could also extract randomness from the multiphoton components of the weak coherent pulse, as long as the photon number is preserved in the channel. When the photon number is preserved, we know that Eve is not performing photon-number-splitting attack and our method allows Alice and Bob to bound the probability that the photon number is preserved, i.e., $q_{n|n}$, thanks to the homodyne detection with a random LO phase.

\section{Conclusion}\label{sec: conclusion}
In conclusion, we have proposed a QKD protocol which shares the features of discrete and continuous variable protocols. A key advantage of our protocol over most existing CV-QKD protocols is that the need for a common phase reference between Alice and Bob is now completely eliminated, which greatly simplifies the system configuration.

We then analyse the security of the protocol under the assumption of collective attacks and in the asymptotic limit. To bound Eve's uncertainty about the key, we adopt the refined Pinsker's inequality proposed in Ref.~\cite{schwonnek2021device} and the channel estimation technique presented in Ref.~\cite{lavie2021estimating}. Our security analysis framework allows us to work in the trusted device scenario which permits us to incorporate characterised device imperfections (such as limited detector efficiency) into our security analysis.

From our simulation, we find that the protocol supports high key generation rate, especially in the low loss regime. For instance, operating at a repetition rate of 1 GHz would amount to an asymptotic secret key rate of about 6.9~Mbit/s at 5~km assuming an effective detector efficiency of 72\%. While its performance is not as robust to loss as compared to DV-QKD protocols, it shows promising potential for applications in metropolitan-distance QKD as it could potentially distribute keys across 15~km with a realistic detector efficiency.

\section*{Acknowledgement}
We would like to thank Xiao Jie Tan for his contribution in studying the direct reconciliation version of this protocol. We would also like to thank Bing Qi, Ren\'e Schwonnek, Wen Yu Kon, and Emilien Lavie for helpful discussion. We acknowledge funding support from the National Research Foundation of Singapore (NRF) Fellowship grant (NRFF11-2019-0001) and NRF Quantum Engineering Programme 1.0 grant (QEP-P2) and the Centre for Quantum Technologies.

\onecolumngrid
\appendix
\newpage

\section{Non-ideal homodyne detector} \label{sec: imperfect detector}
\subsection{POVM element}
In the Section~\ref{sec: security analysis} of the main text, we have assumed that the homodyne detection that we use is an ideal homodyne detector with 100\% quantum efficiency and no additional electronic noise. To account for the imperfection in the homodyne detector, it is sufficient to consider imperfect quantum efficiency detector since electronic noise is equivalent to optical loss after re-calibration of the vacuum noise \cite{appel2007electronic}. Alternatively, one could also use the model presented in Ref.~\cite{lin_trusted_2020}.

Now, to model the imperfect quantum efficiency, we model the realistic homodyne detector using a virtual beam-splitter with transmittivity $\edet$, followed by an ideal quadrature measurement. Suppose the incoming signal mode is associated to the annihilation operator $\ah$ and the other input mode (the vacuum mode) to the beam-splitter is associated to the annihilation operator $\bh$. Let the output mode of the beam-splitter in which the ideal quadrature measurement is performed be denoted by $\ah'$. Then we have
\begin{equation}
    \ah' = \sqrt{\edet} \ \ah + \sqrt{1-\edet} \  \bh
\end{equation}
and
\begin{equation} \label{eq: lossy measurement model}
    \hat{Q}_\varphi^{(a')} = \sqrt{\edet} \ \hat{Q}_\varphi^{(a)} + \sqrt{1-\edet} \ \hat{Q}_\varphi^{(b)},
\end{equation}
where $\hat{Q}_\varphi^{(a_j)} = \ah_j e^{-i\varphi} + \ah_j^\dagger e^{i\varphi}$ is the quadrature operator in mode $\ah_j$ with local oscillator phase $\varphi$. Now, the ideal POVM element $\Pi^\varphi(q)$ can be written as
\begin{equation}
    \Pi^\varphi(q) = \ketbra{q(\varphi)}{q(\varphi)} = \int \diff{q'} \ \delta(q - q') \ \ketbra{q'(\varphi)}{q'(\varphi)},
\end{equation}
where $\delta(\cdot)$ is the Dirac's delta function. Recalling the following identity
\begin{equation}
    \delta(q-q') = \frac{1}{2\pi} \int \diff{\lambda} \ e^{i\lambda(q-q')},
\end{equation}
and then using the spectral decomposition of the quadrature operator, we can write the ideal POVM element as
\begin{equation}
    \Pi^\varphi(q) = \frac{1}{2\pi} \int \diff{\lambda} \ e^{i\lambda (\hat{Q}_\varphi - q)}.
\end{equation}

Applying the above relation to the quadrature measurement in mode $\ah'$, we have
\begin{equation}
    \Pi_{\edet}^\varphi(q) = \frac{1}{2\pi} \int \diff{\lambda} \ e^{i\lambda (\hat{Q}^{(a')}_\varphi - q)} = \frac{1}{2\pi} \int \diff{\lambda} \ e^{i\lambda (\sqrt{\edet} \hat{Q}^{(a)}_\varphi - q)} \bra{v} e^{i \lambda \sqrt{1-\edet} \hat{Q}_\varphi^{(b)}} \ket{v}
\end{equation}
by tracing over the mode $\bh$. After using spectral decomposition again (for quadrature operators in mode $\ah$ and $\bh$), we have
\begin{equation}
    \Pi_{\edet}^\varphi(q) = \frac{1}{2\pi} \int \diff{\lambda} \int \diff{x} \ e^{i\lambda (\sqrt{\edet} q' - q)} \ \ketbra{q'(\varphi)}{q'(\varphi)}  \int \diff{y} \ e^{i \lambda \sqrt{1-\edet} y} \abs{\psi_0(y)}^2  ,
\end{equation}
where the projector $\ketbra{q'(\varphi)}{q'(\varphi)}$ is the quadrature projection for mode $\ah$ and $\psi_0(y)$ is the wavefunction of the vacuum state. Performing the integration over $y$, we have
\begin{equation}
    \int \diff{y} \ e^{i \lambda \sqrt{1-\edet} y} \abs{\psi_0(y)}^2 = \exp \left[ -\frac{\lambda^2 (1-\edet)}{2} \right].
\end{equation}
Now, performing the integration over $\lambda$, we have
\begin{equation}
    \int \diff{\lambda} \ e^{i\lambda (\sqrt{\edet} q' - q)} e^{-\lambda^2 (1-\edet)/2} = \sqrt{\frac{2\pi}{1-\edet}} \exp\left[ -\frac{(\sqrt{\edet}q'- q)^2}{2(1-\edet)} \right].
\end{equation}
Hence, finally we obtain
\begin{equation} \label{eq: realistic POVM}
    \Pi_{\edet}^\varphi(q) = \int \diff{q'} \ \frac{1}{\sqrt{2\pi(1-\edet)}} \exp\left[ -\frac{(q - \sqrt{\edet}q')^2}{2(1-\edet)} \right] \ketbra{q'(\varphi)}{q'(\varphi)},
\end{equation}
i.e., a convolution of the ideal POVM element with a Gaussian function.

\subsection{Block-diagonal structure}
A crucial element of our security proof is the argument that Bob's measurement is block-diagonal in the photon number basis. Here, we will show that the same structure is still preserved when we have imperfect detector.

Given the expression for the POVM element derived in the previous section, we can also obtain the POVM element for the quadrature measurement in the early and late time-bins
\begin{multline}
    \Pi^\varphi_{\edet}(q_0, q_1) = \frac{1}{2\pi(1-\edet)} \int \diff{q_0'} \  \exp\left[ -\frac{(q_0 - \sqrt{\edet}q_0')^2}{2(1-\edet)} \right] \ketbra{q_0'(\varphi)}{q_0'(\varphi)} \otimes \\
    \int \diff{q_1'} \  \exp\left[ -\frac{(q_1 - \sqrt{\edet}q_1')^2}{2(1-\edet)} \right] \ketbra{q_1'(\varphi)}{q_1'(\varphi)}.
\end{multline}
Taking into account phase randomisation, we have
\begin{equation}
    \Pi_{\edet}(q_0, q_1) = \frac{1}{2\pi} \int \diff{\varphi} \  \Pi^\varphi_{\edet}(q_0, q_1).
\end{equation}
To see that the block-diagonal structure is preserved, observe that only the projectors depend on $\varphi$ while the exponential terms do not. As such, we could switch the order of the integration and perform the integration over $\varphi$ before performing the convolutions. However, the integration over $\varphi$ is exactly the same as the one we did in \eqref{eq: block-diagonal measurement} and hence the block-diagonal structure is preserved.

\subsection{Energy measurement}
Our model for imperfect homodyne detector \eqref{eq: lossy measurement model} allows us to relate the observed homodyne measurement $\hat{Q}_\varphi^{(a')}$ to an ideal quadrature measurement in the signal mode $\hat{Q}_\varphi^{(a)}$. If one is interested in the number of photons in the signal mode, we have to calculate the expectation value of the number operator in mode $\ah$, i.e., $\hat{N}^{(a)} = \ah^\dagger \ah$. To that end, consider the number operator in mode $\ah'$
\begin{equation}
\begin{split}
        \hat{N}^{(a')} = \ah'^\dagger \ah' &= (\sqrt{\edet} \ah + \sqrt{1-\edet} \bh)^\dagger (\sqrt{\edet} \ah + \sqrt{1-\edet} \bh) \\
        &= \edet \ah^\dagger \ah + (1-\edet) \bh^\dagger \bh + \sqrt{\edet(1-\edet)} (\ah^\dagger \bh + \ah \bh^\dagger).
\end{split}
\end{equation}
Since the input state in mode $\bh$ is the vacuum state, we have
\begin{equation}
    \mean{\hat{N}^{(a')}} = \edet \mean{\hat{N}^{(a)}}
\end{equation}
and since
\begin{equation}
    \mean{\hat{N}^{(a')}}_{\mu = \mu_i} = \frac{\mean{q_\beta^{\ 2} }_{\mu = \mu_i} - 1}{2},
\end{equation}
we have
\begin{equation}
    \omega_i := \mean{\hat{N}^{(a)}}_{\mu = \mu_i} = \frac{\mean{q_\beta^{\ 2} }_{\mu = \mu_i} - 1}{2 \edet},
\end{equation}
where 
\begin{equation}
    \mean{q_\beta^{\ 2} }_{\mu = \mu_i} = \left\langle \int \frac{\diff{\varphi}}{2\pi} \  \left(\hat{Q}_\varphi^{(a')}\right)^2 \right\rangle_{\mu = \mu_i}
\end{equation}
is the observed mean-square quadrature measurement when Alice chooses intensity $\mu = \mu_i$.

\subsection{Probability distribution of the discretised quadrature measurement}
Finally, the imperfect detection efficiency would affect the probability distribution of the discretised quadrature measurement of Fock states $\ket{m}$. Suppose Bob receives the Fock state $\ket{m}$ in the signal mode $\ah'_\beta$, for each $\beta$, the observed probability distribution would be given by
\begin{equation}
    C^\beta_{\nu | m} = \sum_{k \leq m} \binom{m}{k} (1-\edet)^{m-k} \edet^k C^{\beta,\text{ideal}}_{\nu | k},
\end{equation}
where
\begin{equation}
    C^{\beta,\text{ideal}}_{\nu | k} = \int_{\nu \delta}^{(\nu + 1) \delta} \diff{q} \abs{\psi_k(q)}^2
\end{equation}
is the probability of the outcome of the \textit{ideal} quadrature measurement on the Fock state $\ket{k}$ to be inside interval $I_\nu$.

\section{Postselection map and pinching channel}\label{sec: postselection and pinching}
In this section, we will derive the expression for the postselected state and the pinched state which are needed to apply the refined Pinsker's inequality. To define the postselection map $\ps$ and pinching channel $\pinching$, it is convenient to implement the channels in terms of isometries. Note that all the calculations done in this section are conditioned on Alice preparing $\N = n$ photons as well as her choosing the signal intensity $\mu = \mu_0$.

First, the untrusted quantum channels can be thought of as isometries $\U_{A' \rightarrow BE\M}$ that maps the system $A'$ to $B$ and $E$ where $E$ is held by Eve and $\M$ is the classical register that stores the number of photons that Bob receives
\begin{equation}
    \U_{A' \rightarrow BE \M}: \qquad \ket{a_x}^n_{A'} \rightarrow \sum_{m} \sqrt{q_{m|n}} \ket{\phi_{a,x,n}^m}_{BE} \ket{m}_{\M},
\end{equation}
where $a \in \{0,1\}$, $x \in \{X,Z\}$ and $q_{m|n}$ is the probability that Bob receives $m$ photons conditioned on Alice sending $n$ photons. The state $\{\ket{\phi_{a,x,n}^m}_{BE}\}_{a,x,n,m}$ are entangled states shared by Bob and Eve. In passing, we remark that by tracing out the systems $E$ and $\M$, we should recover the completely-positive and trace-preserving (CPTP) map $\E^{(n)}_{A' \rightarrow B}$

Now, using Naimark's dilation, Bob's measurement can also be thought of as an isometry $\V_{\X BE\M \rightarrow \X R B E\M}$
\begin{equation}
    \V_{\X BE \M \rightarrow \X R B E\M}: \qquad \ket{x}_\X \ket{\phi_{a,x,n}^m}_{BE} \ket{m}_{\M} \rightarrow \ket{x}_\X \sum_{b} \ket{b}_{R}\ket{\tilde{\Omega}_{a,x,n}^{b,m}}_{BE} \ket{m}_{\M} ,
\end{equation}
where the measurement outcome $\B$ can be obtained by performing projective measurements $\{\ketbra{b}{b}\}_{b}$ on the ancilla system $R$. The sub-normalised states $\left\{\ket{\tilde{\Omega}_{a,x,n}^{b,m}}_{BE}\right\}_{a,x,n,m,b}$ are given by
\begin{equation}
    \ket{\tilde{\Omega}_{a,x,n}^{b,m}}_{BE} = \sqrt{M^{(m)}_{b|x}} \otimes \1_E \ket{\phi_{a,x,n}^m}_{BE},
\end{equation}
with $M_{b|x}^{(m)}$ is the POVM element of Bob's measurement in the $m$-photon subspace.

For the postselection process, we consider the isometry $\W_{\X R \rightarrow \X R \Cs}$ where the system $\Cs$ is the classical register that indicates whether the state will be kept or discarded in the postselection process
\begin{equation}
    \W_{\X R \rightarrow \X R \Cs}: \qquad
    \begin{cases}
    \ket{X}_\X \ket{b}_R \rightarrow \ket{X}_\X \ket{b}_R \ket{\discard}_{\Cs} & \forall b\\
    \ket{Z}_\X \ket{b}_R \rightarrow \ket{Z}_\X \ket{b}_R \ket{\discard}_{\Cs} & \text{if } \B \in \{\varnothing, ?\}\\
    \ket{Z}_X \ket{b}_R \rightarrow \ket{Z}_\X \ket{b}_R \ket{\keep}_{\Cs} & \text{if } \B \in \{0,1\}
    \end{cases}.
\end{equation}
Note that applying the isometries and then projecting to the projector $\ketbra{\keep}{\keep}_\Cs$ and $\sum_{m \leq n} \ketbra{m}{m}_\M$ and followed by tracing out the irrelevant systems is equivalent to the postselection map $\ps$. Doing the projection, we get
\begin{equation}
\begin{split}
    \sum_{m \leq n} q_{m|n} p_Z \ketbra{Z}{Z}_\X &\otimes
    \left( \frac{1}{2} \sum_{a,a'} \sum_{b,b' \in \{0,1\}} \ketbra{a\vphantom{'}}{a'}_{A} \otimes 
    \ketbra{b\vphantom{'}}{b'}_R \otimes
    \ketbra{\tilde{\Omega}_{a,Z,n}^{b,m}}{\tilde{\Omega}_{a',Z,n}^{b',m}}_{BE}
    \right)\\
    &\hspace{2cm}\otimes \ketbra{m}{m}_{\M} \otimes \ketbra{\keep}{\keep}_\Cs,
\end{split}
\end{equation}
and tracing out the systems that are either held by Eve or announced during the protocol, namely $E$, $\X$, $\Cs$ and $\M$, the final state is given by
\begin{equation}
    \tilde{\sigma}^{(n)}_{A B R} =\sum_{m \leq n} \sum_{b,b' \in \{0,1\}}
    \left(\1_{A} \otimes \sqrt{M^{(m)}_{b|Z}} \right) \tilde{\rho}_Z^{(m,n)} \left(\1_{A} \otimes \sqrt{M^{(m)}_{b'|Z}} \right) \otimes  \ketbra{b\vphantom{'}}{b'}_R,
\end{equation}
where
\begin{equation}
    \tilde{\rho}_Z^{(m,n)} = \Tr_\X \Big[\left(\ketbra{Z}{Z}_\X \otimes \1_{A B} \right) \tilde{\rho}_{\X A B}^{(m,n)} \left(\ketbra{Z}{Z}_\X \otimes \1_{A B} \right) \Big]
\end{equation}
is the (sub-normalised) state conditioned on $Z$-basis and $m$ photons are received by Bob and Alice prepares $n$ photons. To normalise the final state, we need to divide $\tilde{\sigma}^{(n)}_{A B R}$ by the appropriate factor, namely
\begin{equation}
    p_\postselect^{(n)} = \sum_{m \leq n} p_Z \cdot  q_{m|n} \cdot \Pr[b \in \{0, 1\} | \M = m, \N = n, \X = Z].
\end{equation}

Then, denoting
\begin{equation}
    \sigma^{(n)}_{A B R} =
    \frac{\tilde{\sigma}^{(n)}_{A B R}}{p_{\postselect}^{(n)}}
\end{equation}
as the normalised postselected state, we have
\begin{equation}
    \ps \left[ \rho^{(n)}_{X A B} \right] = \tilde{\sigma}^{(n)}_{A B R} = p_{\postselect}^{(n)} \cdot \sigma^{(n)}_{A B R},
\end{equation}
and since the measurement on $R$ is projective, the pinched state can be obtain by
\begin{equation}
\begin{split}
    \pinching \left[ \sigma^{(n)}_{A B R}  \right] =
    \left(\1_{A B} \otimes \ketbra{0}{0}_R \right)
    & \sigma^{(n)}_{A BR}
    \left(\1_{A B} \otimes \ketbra{0}{0}_R \right)\\
    & + \left(\1_{A B} \otimes \ketbra{1}{1}_R \right)
    \sigma^{(n)}_{A B R}
    \left(\1_{A B} \otimes \ketbra{1}{1}_R \right).
\end{split}
\end{equation}

\bibliographystyle{unsrtnat}
\bibliography{DVCV}
\end{document}